\documentclass[showpacs,showkeys, groupedaddress,twocolumn, floatfix,aps, prb,reprint]{revtex4-1}
\usepackage[english]{babel}
\usepackage{fontenc}
\usepackage{graphicx}
\usepackage{amsmath}
\usepackage{epstopdf}
\usepackage{amssymb}
\usepackage{amsbsy}
\usepackage{amscd}
\usepackage{float}
\usepackage{color}
\usepackage{tikz}
\usepackage{braket}
\usepackage{standalone}
\usepackage{bm}
\usepackage{siunitx}
\usepackage[normalem]{ulem} 
\usepackage{verbatim}
\usepackage{url}
\usetikzlibrary{positioning,calc}
\tikzset{>=latex}

\usepackage[verbose,hypertexnames=false,bookmarksopenlevel=1,filecolor=blue,
linkcolor=blue,citecolor=blue,urlcolor=blue,pdfstartview=FitH,bookmarksopen,bookmarksnumbered,
colorlinks,plainpages=false,linktocpage]{hyperref}

\begin{document}

\title{Many-Body Localization: Transitions in Spin Models}
\author{ John Schliemann$^1$, Jo$\tilde{\rm a}$o Vitor I. Costa$^2$, Paul Wenk$^1$, and J. Carlos Egues$^2$}
\address{$^1$Institute for Theoretical Physics,
  University of Regensburg, Regensburg, Germany\\
  $^2$Instituto de Fısica de Sao Carlos, Universidade de Sao Paulo,
  Sao Carlos, Brazil}

\date{\today}

\begin{abstract}
  We study the transitions between ergodic and many-body localized phases
  in spin systems, subject to quenched disorder, including the Heisenberg chain
  and the central spin model. In both cases systems with
  common spin lengths $1/2$ and $1$ are investigated via exact numerical
  diagonalization and random matrix techniques.
  Particular attention is paid to the sample-to-sample variance
  $(\Delta_sr)^2$ of the averaged consecutive-gap ratio $\langle r\rangle$
  for different disorder realizations. For both types of systems and spin
  lengths we find a maximum in $\Delta_sr$ as a function of disorder strength,
  accompanied by an inflection point of $\langle r\rangle$, signaling the
  transition from ergodicity to many-body localization. The critical
  disorder strength is found to be somewhat smaller than the values reported in
  the recent literature.
  Further information about the transitions can be
  gained from the probability distribution of expectation values within a
  given disorder realization.
\end{abstract}

\maketitle

\section{Introduction}

Many-body localization has become in the last
years one of the most intensively growing areas of research in
condensed matter physics and beyond
\cite{Nandkishore15,Altman15,Imbrie16a,Agarwal17,Luitz17,Haldar17,Abanin17,Abanin18}.
It denotes the absence of thermalization in an
isolated interacting quantum system in the presence of typically strong
disorder. In the opposite ergodic phase the
eigenstate thermalization hypothesis is fulfilled stating that
any appropriate subsystem of the isolated total system (being in a pure state)
is accurately described by equilibrium statistical mechanics
\cite{Deutsch91,Srednicki94}. On the other hand, the presence of interactions
distinguishes many-body localization from traditional Anderson localization
\cite{Anderson58,Evers08}.

In this work we revisit the transition from the ergodic to the many-body
localized phase in disordered Heisenberg spin chains and compare it with the
behavior of central spin models also subject to quenched disorder. For both
types of systems we consider the spin lengths $1/2$ and $1$.
We introduce a novel and very useful tool to quantify the transition,
namely, the sample-to-sample
variance of the averaged consecutive-gap ratio and the underlying probability
distribution. As we will see in the following, a maximum of this
variance signals, for both of the above systems and both spin lengths,
the transition between ergodicity and many-body localization.
This maximum is accompanied by an inflection point of of the averaged
consecutive gap ratio $\langle r\rangle$,
suggesting a close analogy to classic phase transitions with
$\langle r\rangle$ being an order parameter. These central observations
are summarized in Figs.~\ref{chainfig2}, \ref{censpinfig2} for the
Heisenberg chain and the central spin model, respectively.

An additional tool in the analysis of the transition from the ergodic to the
many-body localized phase is the probability distribution of expectation
values within a given disorder realization. The corresponding data
is contained in Figs~\ref{chainfig4} and \ref{censpinfig4}.

This paper is further organized as follows: In section \ref{ModelandApproach} we
introduce the spin models to be studied and summarize the underlying theoretical
techniques. Our numerical results are presented in section
\ref{NumericalResults}, and we close with a summary and an outlook in section
\ref{SummaryandOutlook}.

\section{Model and Approach}
\label{ModelandApproach}

\subsection{Spin Hamiltonian}

We study a periodic Heisenberg spin chain interacting with an additional
central spin and being subject to an uniaxial quenched disorder field on
each site of the chain,
\begin{equation}
  H=J\sum_{i=1}^K\vec I_i\cdot\vec I_{i+1}
  +\frac{A}{K}\vec S\cdot\sum_{i=1}^K\vec I_i+2S\sum_{i=1}^Kh_iI^z_i\,,
  \label{hamiltonian}
\end{equation}
where the parameter $J$ describes the coupling of the $K$ chain (or bath) spins
$\vec I_{i}=\vec I_{i+K}$ to their nearest neighbors. The coupling to the central
spin $\vec S$ is parametrized by $A$, and the random magnetic field $h_i$
is chosen from a uniform distribution within the interval $[-h,h]$.
In what follows all spins will have length $S=I=1/2$ or $S=I=1$.

For $S=I=1/2$, the Hamiltonian (\ref{hamiltonian}) with $A=0$,
i.e. the Heisenberg spin-$1/2$
chain with quenched on-site disorder, is 
a workhorse of numerical studies of many-body localization
\cite{Santos04,Pal10,Badarson12,Luitz15,Chandran15,Agarwal15,Baygan15,Devakul15,Luitz16a,Luitz16b,Geraedts16,Khemani16,Lim16,Serbyn16,Enss17,Acevedo17,Khemani17,Filho17,Geraedts17,Xu18,Doggen18,Suntajs19,Panda20,Chanda20a,Suntajs20,Laflorencie20,Sierant20,Dhara20,Chanda20b,Throckmorton20}. 
The factor $2S$ in front of the disorder term makes contact to the usual
parametrization for $S=1/2$ and ensures an appropriate scaling behavior
of the Hamiltonian for larger spin lengths. Specifically, when considering
all spins as classical vectors of constant length (and not as operators),
the transformation $S\mapsto q S$, $I\mapsto q I$ leads to
$H\mapsto q^2H$. Thus, disorder-induced effects on the dynamics should
occur at the same disorder strength $h$.
As we shall see in section \ref{NumericalResults}, this remains approximately
true when switching between $S=I=1/2$ and $S=I=1$.

Recent work by Hetterich {\rm et al.} \cite{Hetterich18} studied the full
model (\ref{hamiltonian}) for $S=I=1/2$
concentrating on the case $J=1$. As these authors
argue, dividing the interaction parameter $A$ by the number of
central spins $K$, as done in Eq.~(\ref{hamiltonian}), ensures that the
spectral bandwidth of that coupling term is approximately independent of the
system size. We will see in the following that this stipulation has indeed
advantages when comparing data for different numbers of spins.

\subsection{Random Matrix Theory}
\label{RandomMatrixTheory}

An important method to distinguish a many-body localized phase from an ergodic
phase is random matrix theory, which is a theory for statistical
fluctuations of energy levels of a given quantum system
\cite{Guhr98}. A modern tool to analyze the energy level statistics is
the consecutive gap ratio \cite{Oganesyan07} defined as
\begin{equation}
  r_n=\frac{\min\{s_n,s_{n-1}\}}{\max\{s_n,s_{n-1}\}}
  =\min\left\{\bar r_n,\frac{1}{\bar r_n}\right\}
  \quad,\quad\bar r_n=\frac{s_n}{s_{n-1}}\,,
  \label{ratgap1}
\end{equation}
where $s_n=e_{n+1}-e_n$ is the difference between two neighboring energy levels
$e_{n+1}$, $e_n$. In the strictly many-body localized (or integrable)
phase, characterized by an extensive number of independent conserved quantities, the differences $s_n$ obey Poisson
statistics, and the probability distribution for the random variable $r=r_n$
can easily be determined to be \cite{Oganesyan07}
\begin{equation}
  p(r)=\frac{2}{(1+r)^2}\,.
  \label{ratgap2}
\end{equation}
On the other hand, in the fully ergodic phase, a system of the type
(\ref{hamiltonian}) is generally assumed to be described by the
Gaussian orthogonal ensemble (GOE) of random matrices \cite{Guhr98}.
Here the analysis of small random matrices mimicking the Hamiltonian predicts
the pertaining probability distribution to be \cite{Atas13}
\begin{equation}
  p(r)=\frac{27}{4}\frac{r+r^2}{(1+r+r^2)^{5/2}}\,,
  \label{ratgap3}
\end{equation}
which can be seen as an analog of the classic Wigner surmises for
probability distributions governing the traditional random variable $s=s_n$
\cite{Guhr98}.

As a consequence, the lowest moments of the probability distribution
(\ref{ratgap2}) in the integrable case are given by
\begin{eqnarray}
  \langle r\rangle_p & = & 2\ln 2-1\approx 0.3863\,,
  \label{ratgap4}\\
  \langle r^2\rangle_p & = & 3-4\ln 2\approx 0.2274\,,
  \label{ratgap5}\\
  \Delta_p r
  & = & \sqrt{\langle r^2\rangle_p-\langle r\rangle^2_p}\approx 0.2796\,,
  \label{ratgap6}
\end{eqnarray}
with
\begin{equation}
  \langle\cdot\rangle_p=\int_0^1dr(p(r)\cdot)\,,
  \label{ratgap7}
\end{equation}  
whereas in the ergodic situation (\ref{ratgap3}) we have
\begin{eqnarray}
  \langle r\rangle_p & = & 4-2\sqrt{3}\approx 0.5359\,,
  \label{ratgap8}\\
  \langle r^2\rangle_p & = & \frac{27}{4}\ln\left(1+\frac{2}{\sqrt{3}}\right)
  -\frac{1}{2}-\frac{5}{2}\sqrt{3}\approx 0.3515\,,
  \label{ratgap9}\\
  \Delta_p r & \approx & 0.2536\,,
  \label{ratgap10}
\end{eqnarray}

\subsection{Statistical Data Analysis}
\label{StatisticalDataAnalysis}

\subsubsection{Disorder ensemble}

Consider an ensemble of $Q$ realizations of the local disorder field
$h_i$, $i\in\{1,\dots,K\}$.
Each disorder realization, or sample, labeled by $\alpha\in\{1\dots,Q\}$
leads to a probability
distribution $p_{\alpha}(r)$ for the consecutive gap ratio $r\in[0,1]$.
Given an arbitrary function $f(r)$, these distributions determine the
realization-dependent averages
(or expectation values)
\begin{equation}
  \langle f\rangle_{\alpha}=\int_0^1drp_{\alpha}(r)f(r)
  \label{stat1}
\end{equation}
with variances
\begin{equation}
  (\Delta_{\alpha}f)^2=\langle f^2\rangle_{\alpha}-\langle f\rangle_{\alpha}^2\,.
  \label{stat2}
\end{equation}
The disorder-averaged probability distribution $p(r)$ {\em at given disorder
strength} (and other system parameters) is
\begin{equation}
  p(r)=\lim_{Q\to\infty}\frac{1}{Q}\sum_{\alpha=1}^Qp_{\alpha}(r)\,,
  \label{stat3}
\end{equation}
and a good numerical estimate for this quantity is
\begin{equation}
  p(r)\approx\frac{1}{Q}\sum_{\alpha=1}^Qp_{\alpha}(r)\quad,\quad Q\gg 1\,,
  \label{stat4}
\end{equation}
for sufficiently large $Q$. 
The disorder-averaged expectation values of $f(r)$ reads
\begin{equation}
  \langle f\rangle_p=\int_0^1drp(r)f(r)
  =\lim_{Q\to\infty}\frac{1}{Q}\sum_{\alpha=1}^Q\langle f\rangle_{\alpha}\,.
  \label{stat5}
\end{equation}

\subsubsection{Probability distribution for average within sample}

On the other hand, we can view the numbers $x=\langle f\rangle_{\alpha}$
as random variables according to the distribution
\begin{eqnarray}
  s(x) & = & \frac{1}{(2h)^K}\int_{-h}^hdh_1\cdots\int_{-h}^hdh_K\nonumber\\
  & & \qquad\cdot\delta\left(x-\int_0^1drp(r;h_1,\dots,h_K)f(r)\right)
  \label{stat6}
\end{eqnarray}
where $p(r;h_1,\dots,h_K)=p_{\alpha}$ is the probability distribution
within a system with local disorder fields $h_1,\dots,h_K$ forming the
disorder realization $\alpha$.
Thus, the disorder-averaged expectation
value of $f(r)$ can be formulated as
\begin{equation}
  \langle f\rangle_s=\int dxs(x)x
  =\lim_{Q\to\infty}\frac{1}{Q}\sum_{\alpha=1}^Q\langle f\rangle_{\alpha}
  =\langle f\rangle_p\,,
  \label{stat7}
\end{equation}
where the integration goes over all values of $x=f(r)$ for $r\in[0,1]$.
The distribution $s(x)$ does in general not coincide with $p(r)$ even
for $x=f(r)=r$. Moreover $s(x)$ will of course have a dependence on the function
$f$, which we, however, shall suppress in the notation.

The finite average
\begin{equation}
  \bar f=\frac{1}{Q}\sum_{\alpha=1}^Q\langle f\rangle_{\alpha}
  \approx\langle f\rangle_s\,,
  \label{stat8}
\end{equation}
is, again for appropriately large $Q$, an approximation to the
expression (\ref{stat7}). On the other hand, it is a sum of stochastically
independent (and therefore uncorrelated) random variables with identical
probability distribution, so that the resulting joint distribution is
\begin{equation}
  \pi(x_1,\dots,x_Q)=\prod_{\alpha=1}^Qs(x_{\alpha})\,.
  \label{stat9}
\end{equation}
Thus, the expectation value of $\bar f$ is
\begin{equation}
  \langle\bar f\rangle_{\pi}=\frac{1}{Q}\sum_{\alpha=1}^Q\langle f\rangle_s
  =\langle f\rangle_s\,.
  \label{stat10}
\end{equation}

\subsubsection{Sample-to-sample variance}

For the variance pertaining to the expectation value (\ref{stat10}) one finds
\begin{eqnarray}
  & & \left\langle\left(\bar f-\langle f\rangle_s\right)^2\right\rangle_{\pi}
  =\sum_{\alpha,\beta=1}^Q
  \frac{\left\langle\left(\langle f\rangle_{\alpha}-\langle f\rangle_s\right)
    \left(\langle f\rangle_{\beta}-\langle f\rangle_s\right)\right\rangle_{\pi}}
       {Q^2}
    \nonumber\\
    & & \qquad=\frac{1}{Q^2}\sum_{\alpha=1}^Q
  \left\langle\left(\langle f\rangle_{\alpha}-\langle f\rangle_s\right)^2
  \right\rangle_s
  =\frac{(\Delta_sf)^2}{Q}
  \label{stat11}
\end{eqnarray}
with
\begin{eqnarray}
  (\Delta_sf)^2 & = & \langle(f-\langle f\rangle_s)^2\rangle_s\nonumber\\
  & = & \lim_{Q\to\infty}\frac{1}{Q}\sum_{\alpha=1}^Q
  (\langle f\rangle_{\alpha}-\langle f\rangle_s)^2\,.
  \label{stat12}
\end{eqnarray}
Therefore, the fluctuations of the finite average (\ref{stat7}) around
its expectation value (\ref{stat10}) are characterized by the standard
deviation
\begin{equation}
  \Delta_{\pi}\bar f
  =\sqrt{\left\langle\left(\bar f-\langle f\rangle_s\right)^2\right\rangle_{\pi}}
  =\frac{\Delta_sf}{\sqrt{Q}}
  \label{stat13}
\end{equation}
and shows the familiar decay $\propto 1/\sqrt{Q}$.
This result follows of course also from the Lindeberg-Levy central limit
theorem applied to the sum (\ref{stat8}) of random variables.
An approximate expression for the variance (\ref{stat12}) is
\begin{equation}
  (\Delta_sf)^2\approx\frac{1}{Q}\sum_{\alpha=1}^Q
  (\langle f\rangle_{\alpha}-\bar f)^2
  \quad,\quad Q\gg 1\,.
  \label{stat14}
\end{equation}

Note that the variance (\ref{stat12}) also occurs as a contribution to
the variance of $f(r)$ calculated from the disorder-averaged probability
distribution (\ref{stat3}),
\begin{eqnarray}
  (\Delta_pf)^2 & = & \int_0^1drp(r)(f(r)-\langle f\rangle_p)^2
  \label{stat15}\\
  & = & \lim_{Q\to\infty}\frac{1}{Q}\sum_{\alpha=1}^Q
  \int_0^1drp_{\alpha}(r)(f(r)-\langle f\rangle_p)^2\nonumber\\
  & = & \lim_{Q\to\infty}\frac{1}{Q}\sum_{\alpha=1}^Q(\langle f^2\rangle_{\alpha}
  -\langle f\rangle_{\alpha}^2)\nonumber\\
  & & \qquad
  +\lim_{Q\to\infty}\frac{1}{Q}\sum_{\alpha=1}^Q(\langle f\rangle_{\alpha}^2
  -\langle f\rangle_p^2)\nonumber\\
  & = & \lim_{Q\to\infty}\frac{1}{Q}\sum_{\alpha=1}^Q(\Delta_{\alpha}f)^2
  +(\Delta_sf)^2\,.
  \label{stat16}
\end{eqnarray}
Hence, the variance (\ref{stat15}) with respect to the averaged
probability distribution (\ref{stat3}) is the average of all variances
within the disorder realizations around their individual expectation value
of $f$, plus the variance (\ref{stat12}) describing the fluctuations
of these expectation values around their mean. An approximate
expression for the above result is
\begin{eqnarray}
  (\Delta_pf)^2 & \approx & \frac{1}{Q}\sum_{\alpha=1}^Q
  \int_0^1drp_{\alpha}(r)(f(r)-\bar f)^2
   \label{stat17}\\
  & = & \frac{1}{Q}\sum_{\alpha=1}^Q(\Delta_{\alpha}f)^2
  +\frac{1}{Q}\sum_{\alpha=1}^Q
  (\langle f\rangle_{\alpha}-\bar f)^2
  \label{stat18}
\end{eqnarray}
where again $Q\gg 1$.

\subsubsection{Variance of the variance}

Let us  now analyze the statistical fluctuations of the r.h.s. of the
approximate quantity (\ref{stat14}) for finite $Q$ and define
\begin{equation}
  \bar g:=\frac{1}{Q}\sum_{\alpha=1}^Q(\langle f\rangle_{\alpha}-\bar f)^2
  =\frac{1}{Q}\sum_{\alpha=1}^Q
  \left(\langle f\rangle^2_{\alpha}-\bar f^2\right)\,.
  \label{stat19}
\end{equation}
An important difference between the above expression and the quantity
(\ref{stat8}) is that above the summands depend, again via Eq.~(\ref{stat8}),
on all expectation values $\langle f\rangle_{\alpha}$, $\alpha\in\{1,\dots,Q\}$,
and are therefore distributed according to the joint
probability distribution (\ref{stat9}).
The expectation value of the random variable (\ref{stat19}) with respect to
the latter distribution is
\begin{eqnarray}
  \langle\bar g\rangle_{\pi} & = & \frac{1}{Q}\left(
  \sum_{\alpha=1}^Q\left\langle\langle f\rangle^2_{\alpha}\right\rangle_{\pi}
  -\frac{1}{Q}\sum_{\alpha,\beta=1}^Q
  \left\langle\langle f\rangle_{\alpha}\langle f\rangle_{\beta}\right\rangle_{\pi}
  \right)\nonumber\\
  & = & \frac{Q-1}{Q}\left(\langle f^2\rangle_s-\langle f\rangle^2_s\right)
  =\frac{Q-1}{Q}(\Delta_sf)^2
  \label{stat20}
\end{eqnarray}
which approaches the variance (\ref{stat12}) for large $Q$.
Note also that, in contrast to
Eq.~(\ref{stat7}), it holds
\begin{equation}
  \langle f^2\rangle_s=\int dxs(x)x^2
  =\lim_{Q\to\infty}\frac{1}{Q}\sum_{\alpha=1}^Q\langle f\rangle^2_{\alpha}
  \neq \langle f^2\rangle_p\,.
  \label{stat23}
\end{equation}
It is now straightforward to establish that
\begin{eqnarray}
  & & \left\langle(\langle f\rangle^2_{\alpha}-\bar f^2
  -\langle\bar g\rangle_{\pi})
  (\langle f\rangle^2_{\beta}-\bar f^2
  -\langle\bar g\rangle_{\pi})\right\rangle_{\pi}
  \nonumber\\
  & & \qquad\qquad
  =\delta_{\alpha\beta}\left(\langle f^4\rangle_s-\langle f^2\rangle^2_s\right)
  +{\cal O}\left(\frac{1}{Q}\right)\,,
   \label{stat24}
\end{eqnarray}
and therefore, analogously as in Eq.~(\ref{stat11}),
\begin{eqnarray}
  & & \left(\Delta_{\pi}\bar g\right)^2\nonumber\\
  & & \quad=\sum_{\alpha,\beta=1}^Q\frac{\left\langle
  (\langle f\rangle^2_{\alpha}-\bar f^2-\langle\bar g\rangle_{\pi})
    (\langle f\rangle^2_{\beta}-\bar f^2
    -\langle\bar g\rangle_{\pi})\right\rangle_{\pi}}
  {Q^2}
  \nonumber\\
  & & \quad=\frac{(\Delta_sf^2)^2}{Q}+{\cal O}\left(\frac{1}{Q^2}\right)
  \label{stat25}
\end{eqnarray}
where
\begin{eqnarray}
  (\Delta_sf^2)^2 & = & \langle f^4\rangle_s-\langle f^2\rangle^2_s
  =\left\langle\left(f^2-\langle f^2\rangle_s\right)^2\right\rangle_s\nonumber\\
  & = & \left(\Delta_s\left((\Delta_sf)^2\right)\right)^2\,.
  \label{stat26}
\end{eqnarray}

\subsubsection{Consecutive gap ratio}

In what follows we will be mainly concerned with the function $f(r)=r$ where we
have
\begin{eqnarray}
  (\Delta_sr)^2 & = & \lim_{Q\to\infty}\frac{1}{Q}\sum_{\alpha=1}^Q
  (\langle r\rangle_{\alpha}-\langle r\rangle_s)^2\nonumber\\
  & = & \lim_{Q\to\infty}\frac{1}{Q}\sum_{\alpha=1}^Q
  \left(\langle r\rangle^2_{\alpha}-\langle r\rangle^2_s\right)
  \label{stat27}
\end{eqnarray}
with
\begin{equation}
  \langle r\rangle_s=\int_0^1dxs(x)x
  =\lim_{Q\to\infty}\frac{1}{Q}\sum_{\alpha=1}^Q\langle r\rangle_{\alpha}
  =\langle r\rangle_p\,,
  \label{stat28}
\end{equation}
where $s(x)$ is, in accordance with Eq.~(\ref{stat6}), the probability
distribution for the random variable $x=\langle r\rangle_{\alpha}$.
Finally, as seen in Eq.~(\ref{stat26}), the variance of the
variance (\ref{stat27}) is determined by
\begin{eqnarray}
  (\Delta_sr^2)^2 & = & \langle r^4\rangle_s-\langle r^2\rangle^2_s
  =\left\langle\left(r^2-\langle r^2\rangle_s\right)^2\right\rangle_s\nonumber\\
  & = & \left(\Delta_s\left((\Delta_sr)^2\right)\right)^2\,.
  \label{stat29}
\end{eqnarray}

\section{Numerical Results}
\label{NumericalResults}

The Hamiltonian (\ref{hamiltonian}) obviously conserves the $z$-component
of the total spin,
\begin{equation}
  \vec J=\vec S+\sum_{i=1}^K\vec I\quad,\quad
  \left[H,J^z\right]=0\,.
  \label{phases1}
\end{equation}
Thus, in order to apply random matrix theory, the spectra of each invariant
subspace of $J^z$ have to be analyzed separately \cite{Guhr98}.
In this section we present accumulated exact-diagonalization data
from a separate evaluation of all subspaces of $J^z$ except for the four
subspaces of smallest dimension where $|J^z|$ is maximal or differs from
its maximal value by $1$. The number of disorder realizations varies, depending
on system size, between several hundreds and $2\cdot 10^5$.

\subsection{Heisenberg Chain}
\label{HeisenbergChain}

\begin{figure*}
  \includegraphics[width=0.8\columnwidth]{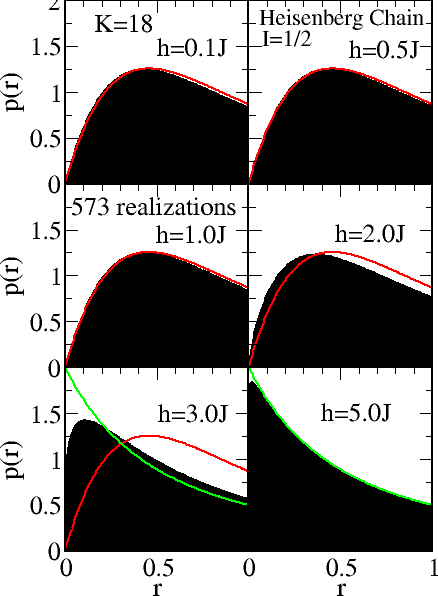}
  \includegraphics[width=0.8\columnwidth]{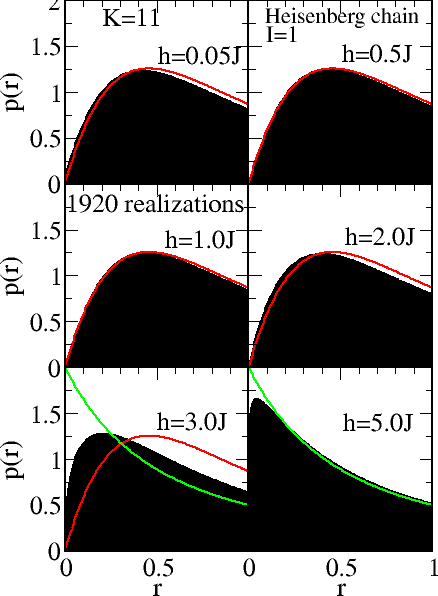}
  \caption{The probability distribution (\ref{stat4}) for the
    consecutive gap ration $r$ of a Heisenberg chain ($A=0$) of $K=18$ spins
    of length $I=1/2$ (left) and of $K=11$ spins of length $I=1$ (right)
    for different disorder strength $h$ obtained from exact-diagonalization
    data. At small disorder the system is ergodic and well described by the
    distribution (\ref{ratgap3}) (red), while with increasing $h$ a
    transition to the Poisson-typed distribution (\ref{ratgap2}) (green) sets
    in.}
  \label{chainfig1}
\end{figure*}
For $A=0$ the central spin $\vec S$ becomes obsolete, and for $I=1/2$ and 
vanishing disorder $h=0$ the resulting Heisenberg chain is integrable via
the Bethe ansatz \cite{Bethe31}. However, this is a rather isolated
point in the phase diagram as seen in Fig.~\ref{chainfig1} showing the
disorder-averaged
probability distribution (\ref{stat4}) obtained from exact-diagonalization
data of Heisenberg chain with spin lengths $I=1/2$ and $I=1$ at different
disorder strengths. For even small 
disorder such as $h=0.1J$ the system shows ergodic statistics (\ref{ratgap3})
while upon increasing $h$ it changes to the Poisson-type distribution
(\ref{ratgap2}). This transition occurs for both spin lengths at about
the same disorder strength, which is a consequence of the scaling factor
$2S=2I$ in the disorder term of the Hamiltonian (\ref{hamiltonian}).
The fact that both spin lengths show such a similar behavior is good news for
semiclassical approaches to many-body localization in spin chains
\cite{Acevedo17,Craps20}.
\begin{figure*}
  \includegraphics[width=1.0\columnwidth]{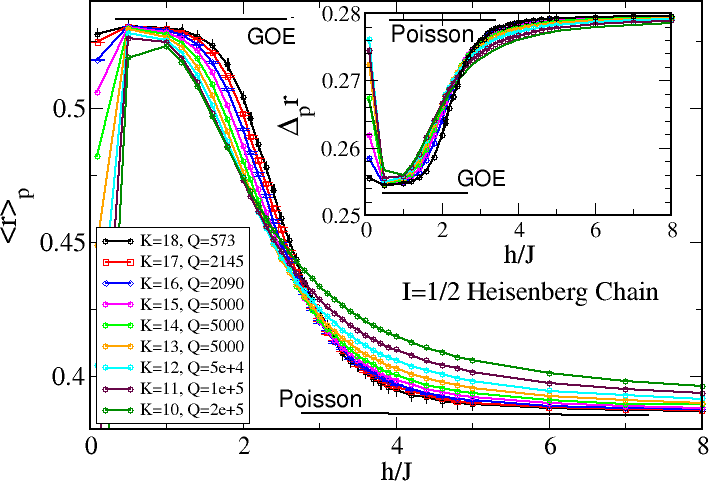}
  \includegraphics[width=1.0\columnwidth]{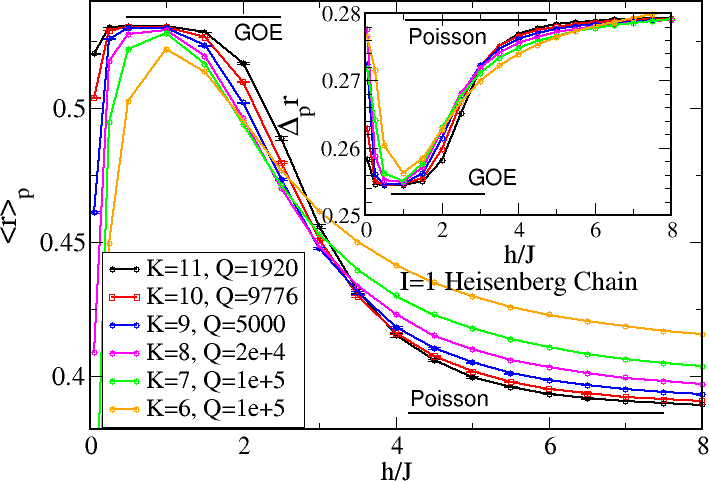}
  \includegraphics[width=1.0\columnwidth]{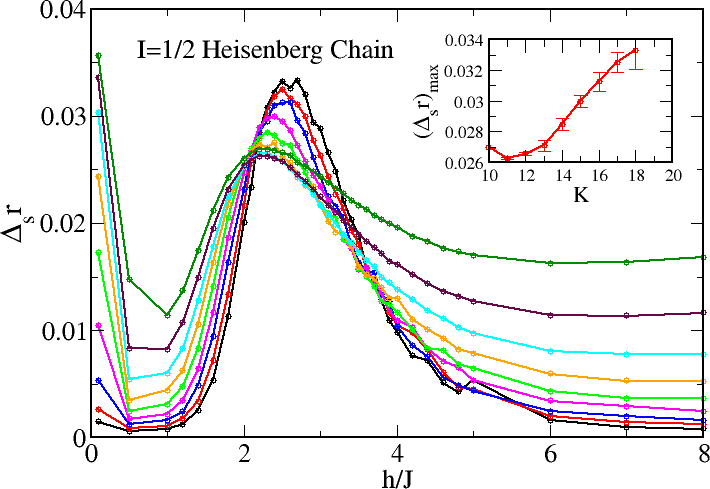}
  \includegraphics[width=1.0\columnwidth]{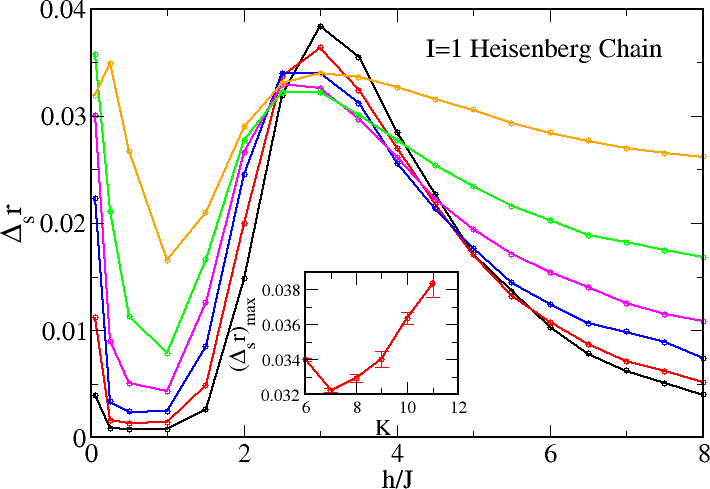}
  \caption{Top panels: The expectation value $\langle r\rangle_p$ as
    a function of disorder strength in Heisenberg chains of spin
    length $I=1/2$ (left) and $I=1$ (right)
    for various system sizes and pertaining numbers of disorder realizations.
    The error bars are determined by Eqs.~(\ref{stat13}), (\ref{stat14}). 
    Inset: The standard deviation $\Delta_pr$ as a function of disorder
    strength. The horizontal lines indicate the expected values
    for GOE and Poissonian statistics as given in
    Eqs.~(\ref{ratgap4})-(\ref{ratgap10}).\\
    Bottom panels: The standard deviation $\Delta_sr$ according to
    Eq.~(\ref{stat14}) for the same parameters as in the top panels.
    For a more detailed view on the data at large system sizes see
    also Fig.~\ref{chainfig3}.
    The insets show the maximum of the standard deviation as a function of
    system size where the error bars follow Eqs.~(\ref{stat26}),
    (\ref{stat29}).}
  \label{chainfig2}
\end{figure*}

\begin{figure}
  \includegraphics[width=1.0\columnwidth]{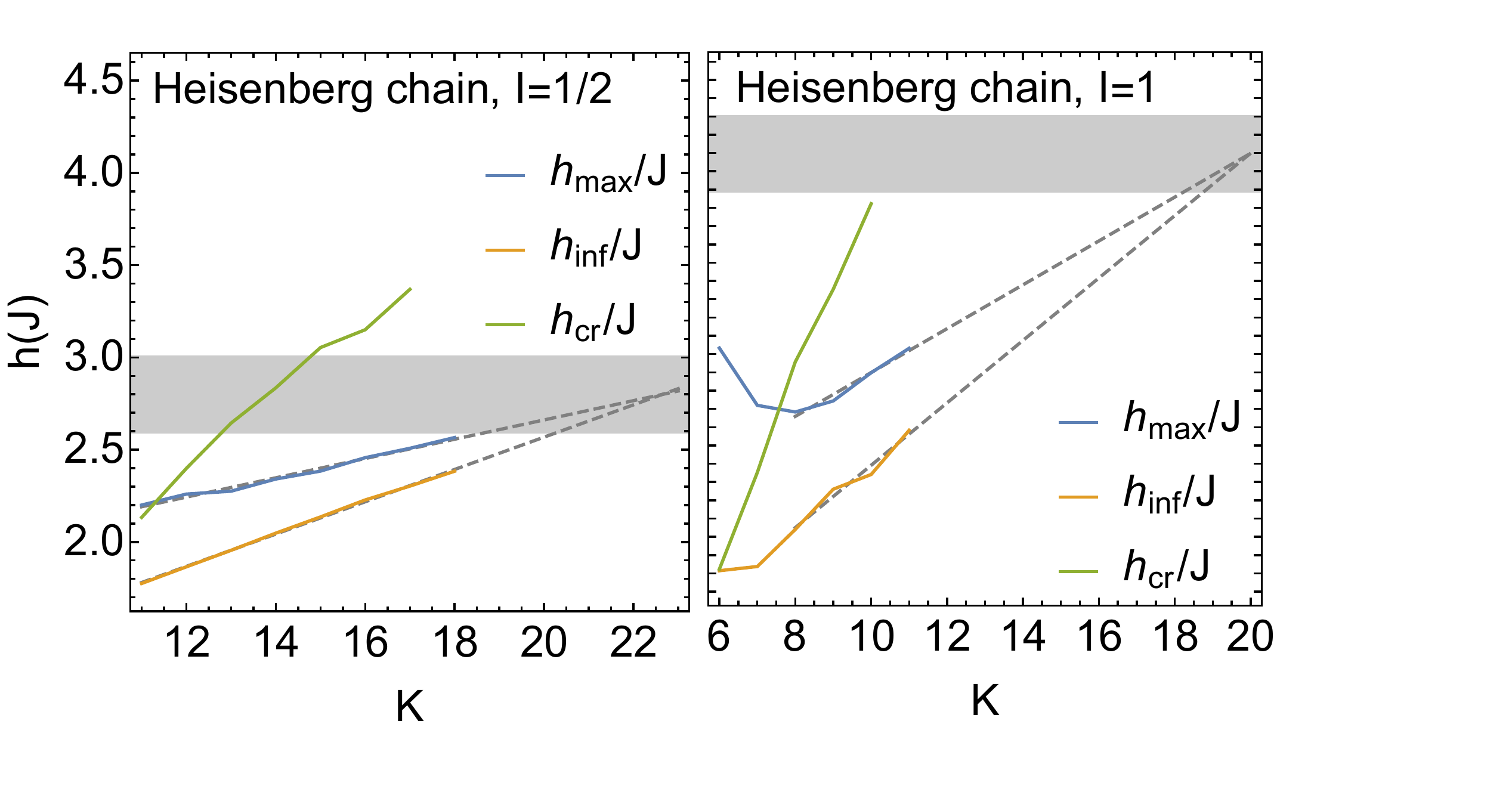}
  \caption{Finite-size transition data for the Heisenberg chain of
    spin length $I=1/2$ (left) and $I=1$ (right). The panels show as a function
    of system size $K$ the position $h=h_{\rm max}$ of $(\Delta_sr)_{\rm max}$,
    the position $h=h_{\rm inf}$ of the inflection point of $\langle r\rangle_s$,
    and the crossing point $h=h_{\rm cr}(K)$ of two curves of the latter
    quantity with consecutive systems sizes $K$ and $K+1$. The dashed lines
    are linear fits to $h_{\rm max}$ and $h_{\rm inf}$, and the shaded regions
    estimate the transition where $h_{\rm max}\approx h_{\rm inf}$}
  \label{chainfig2a}
\end{figure}

In Fig.~\ref{chainfig2} we show the expectation
value $\langle r\rangle_p=\langle r\rangle_s$ as a function of disorder strength
for Heisenberg chains of different sizes along with
the standard deviation $\Delta_pr$ (top panels). The data shows a transition
between the ergodic phase at small $h$ characterized by Eqs.~(\ref{ratgap8}),
(\ref{ratgap10}) to the values (\ref{ratgap4}), (\ref{ratgap6})
of the many-body localized phase. The bottom panels
display the sample-to-sample standard deviation $\Delta_sr$ according to
Eqs.~(\ref{stat14}), (\ref{stat27}).
As seen from the figures, $\Delta_sr$ amounts only to about ten percent
of $\Delta_pr$, which demonstrates via Eq.~(\ref{stat16}) that
$(\Delta_sr)^2$ is only a tiny contribution to the variance $(\Delta_pr)^2$.
On the other hand, $\Delta_sr$ shows a pronounced maximum
$(\Delta_sr)_{\rm max}$ which grows rapidly with system size, as displayed in
the insets of the lower panels. Moreover, in close vicinity to the
corresponding position $h=h_{\rm max}$, $\langle r\rangle_s$. has an inflection
point at $h=h_{\rm inf}$. In Fig.~\ref{chainfig2a} we have
plotted  both disorder strengths for $I\in\{1/2,1\}$
as functions of systems size
$K$, which shows that both quantities seem to converge to a common
value for large $K$.
Thus, the expectation value $\langle r\rangle_s$ and the standard deviation
$\Delta_sr$ show as a function of disorder strength typical features of a
phase transition with the former quantity playing the role of an order
parameter. 

The crossing points of the data shown in the top panels of
Fig.~\ref{chainfig2} are also
often considered as indications for a phase transition.
Therefore, following Refs.~\cite{Pal10,Tikhonov16}, we also plot
in Fig.~\ref{chainfig2a} the positions $h=h_{\rm cr}(K)$ where
two curves of $\langle r\rangle_s$ with consecutive system sizes $K$ and $K+1$
cross. This data set clearly deviates from $h_{\rm max}$, $h_{\rm inf}$
and grows to larger disorder strengths, an observation known as the
``drifting of the critical disorder strength'' with system site
\cite{Pal10,Enss17}. It is an interesting speculation whether
$h_{\rm max}\approx h_{\rm inf}$ and $h_{\rm cr}$ correspond, for large systems,
to two distinct transitions occurring in the same systems.

For the finite-size data depicted in Fig.~\ref{chainfig2a} we estimate 
the transition point to
$h_{\rm max}\approx h_{\rm inf}\approx 2.6J\dots 3.0J$ for spin length
$I=1/2$, and $h_{\rm max}\approx h_{\rm inf}\approx 4.0J\dots 4.5J$ for $I=1$.
These values for the critical disorder strength for the transition from
the ergodic to the many-body lpcalized phase are somewhat smaller than the ones
reported in other works \cite{Pal10,Luitz15,Devakul15,Doggen18,Chanda20a},
which favor, for $I=1/2$, values of $h/J\approx 4$ or larger. However, some
of these works \cite{Pal10,Luitz15,Doggen18} concentrate on chains with an
even number of spins and the subspace with total spin $J^z=0$,
whereas here we also take into account odd numbers
of spins and all subspaces except for those with $|J^z|\in\{IK,IK-1\}$.
Moreover, we introduce a new and different criterion to locate the transition
given by the position of $(\Delta_sr)_{\rm max}$.
\begin{figure}
  \includegraphics[width=0.8\columnwidth]{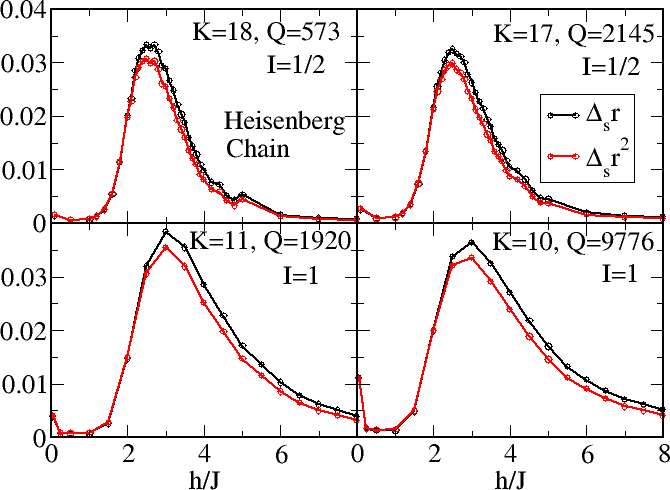}
  \caption{The sample-to-sample standard deviation $\Delta_sr$ along with the
    square root of ``variance of the variance'' (\ref{stat29}) as a function of
    disorder strength $h$ for Heisenberg chains of spin length $I=1/2$
    (top panels) and $I=1$ (bottom panels).
    The data sets are remarkably close to each other, in particular at small
    disorder strength $h$}
  \label{chainfig3}
\end{figure}

\begin{figure*}
  \includegraphics[width=0.8\columnwidth]{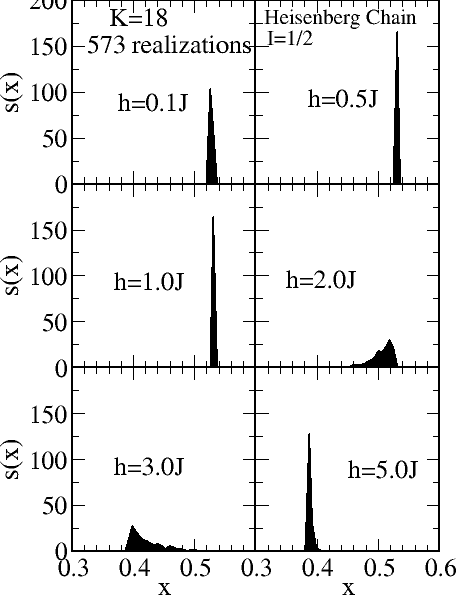}
  \includegraphics[width=0.8\columnwidth]{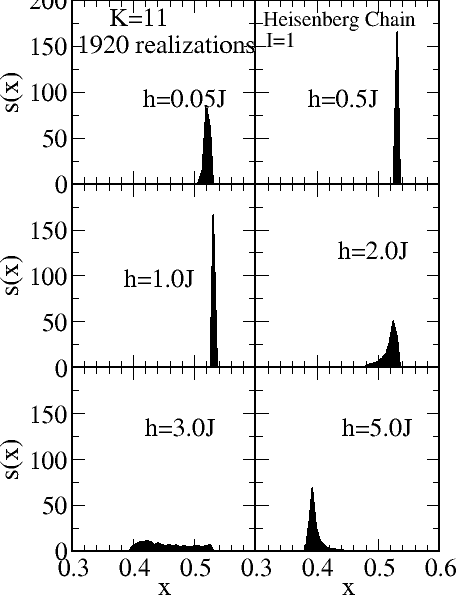}
  \caption{The probability distribution (\ref{stat6}) for the
    realization-specific average $\langle r\rangle_{\alpha}$
    for disordered Heisenberg chains of spin length $I=1/2$ (left) and
    $I=1$ (right).
    The disorder strengths $h$ in both panels are the same as in
    Fig.~\ref{chainfig1}.}
  \label{chainfig4}
\end{figure*}

Fig.~\ref{chainfig3} shows the standard deviation $\Delta_sr$ along with the
square root of ``variance of the variance'' (\ref{stat29})
as a function of disorder
strength for both spin lengths $I=1/2$ and $I=1$.
Remarkably, both quantities are very close to each other, especially at small
disorder  strength $h$. This observation should be taken as an indication that
the underlying probability distribution $s(x)$ is rather narrow since both
quantities become strictly equal, $\Delta_sr=\Delta_sr^2=0$, for
a $\delta$-type distribution.
This conjecture is confirmed by the data of Fig.~\ref{chainfig4}
which displays the probability distribution
(\ref{stat6}) for the realization-specific average $\langle r\rangle_{\alpha}$
with the disorder strengths being the same as in
Fig.~\ref{chainfig1}. The probability distribution is much
narrower than the distribution $p(r)$ and
broadens significantly in the transition region. The latter result is similar
to an obsevation by Pal and Huse \cite{Pal10} who found near the transition
a maximum in the width of the probability distribution of a long-ranged
spin correlator.

We also note that a closer analysis of the data of
the lower panels of Fig.~\ref{chainfig3} suggests that $\Delta_sr$
extrapolates to zero for $K\to\infty$ deep in the ergodic phase
($h/J\approx 1$) as well as deep in the many-body localized phase
($h/J\gtrsim 6$). This would mean that the probability distribution
(\ref{stat6}) would develop into a $\delta$-function in this limit and the
above range of disorder strengths, which is consistent with the data of
Fig.~\ref{chainfig4}. Note that the statement $\lim_{K\to\infty}\Delta_sr=0$
implies, according to Eq.~(\ref{stat16}), that the variance $(\Delta_pr)^2)$
is entirely given by the averaged variances within the individual disorder
realizations.

\subsection{Central Spin Model}
\label{CentralSpinModel}

For $J=0$ the central spin model resulting from the Hamiltonian
(\ref{hamiltonian}) is also integrable via an appropriate Bethe ansatz
and known as the Gaudin model \cite{Gaudin76,Erbe10}. Compared to the Heisenberg
chain, the coupling to the central spin provides an alternative mechanism of
introducing interaction among the bath spins, which are subject to a random
magnetic field.

\begin{figure*}
  \includegraphics[width=0.8\columnwidth]{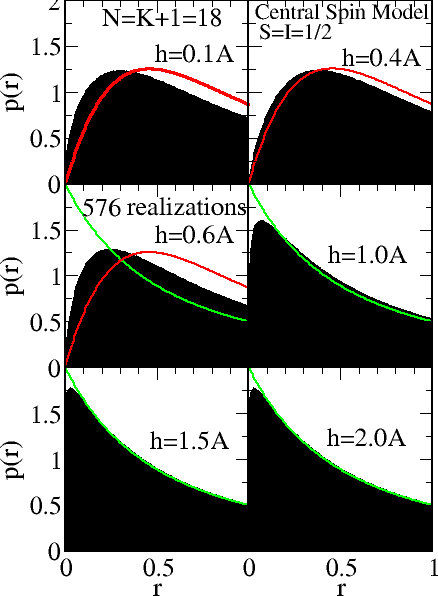}
  \includegraphics[width=0.8\columnwidth]{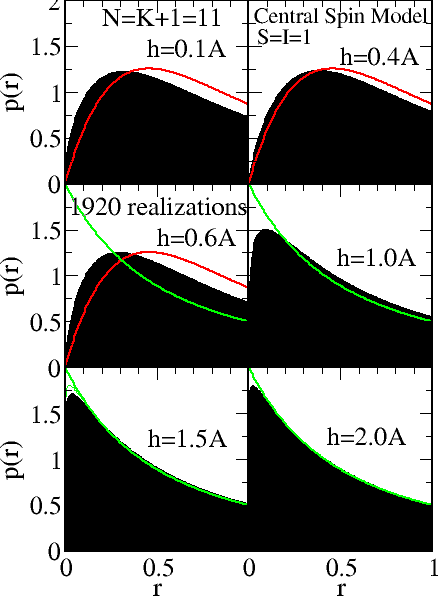} 
  \caption{The probability distribution (\ref{stat4}) for the
    consecutive gap ration $r$ of a central spin model ($J=0$) of $N=K+1=11$
    spins of length $S=I=1/2$ (left) and of $N=K+1=10$ spin of length
    $S=I=1$ (right) for different disorder strength $h$.
    Similarly as for the Heisenberg chain, the data follows for small but finite
    disorder approximately
    the GOE distribution (\ref{ratgap3}) (red), while with increasing  $h$ a
    transition to the Poisson-typed distribution (\ref{ratgap2}) (green) sets
    in.}
  \label{censpinfig1}
\end{figure*}

Fig.~\ref{censpinfig1}
shows data analogous to Fig.~\ref{chainfig1} now for central spin models of
spin length $S=I=1/2$ and $S=I=1$. For small
disorder, the system clearly deviates from the Poisson-type distribution
(\ref{ratgap2}) and shows level repulsion. However, differently from the
case of the Heisenberg chain, the level statistics do not fully
reach the Gaussian orthogonal ensemble but change back before to an integrable
or many-body localized phase.

\begin{figure*}
  \includegraphics[width=1.0\columnwidth]{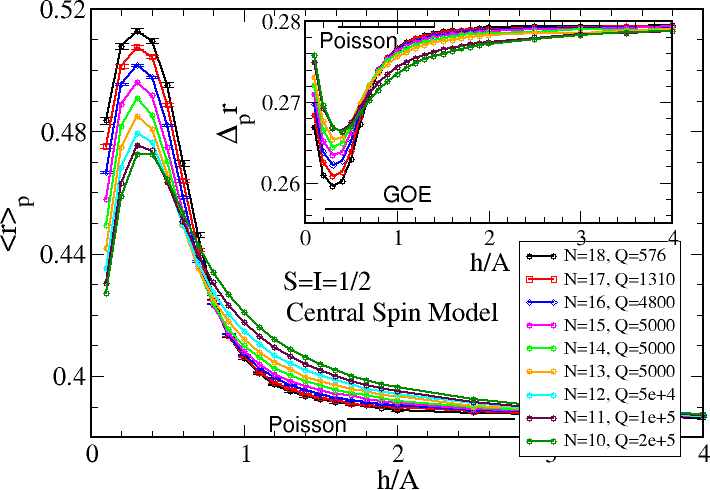}
  \includegraphics[width=1.0\columnwidth]{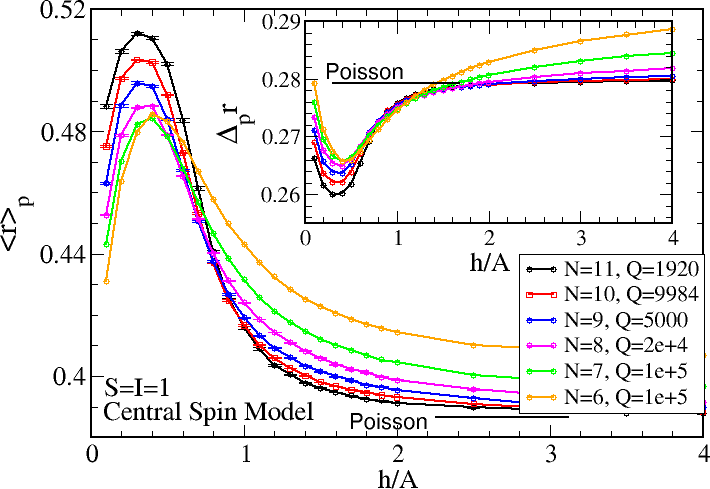}
  \includegraphics[width=1.0\columnwidth]{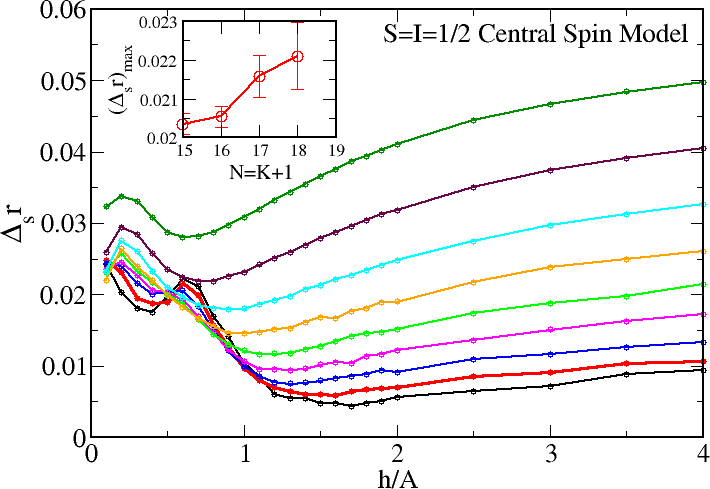}
  \includegraphics[width=1.0\columnwidth]{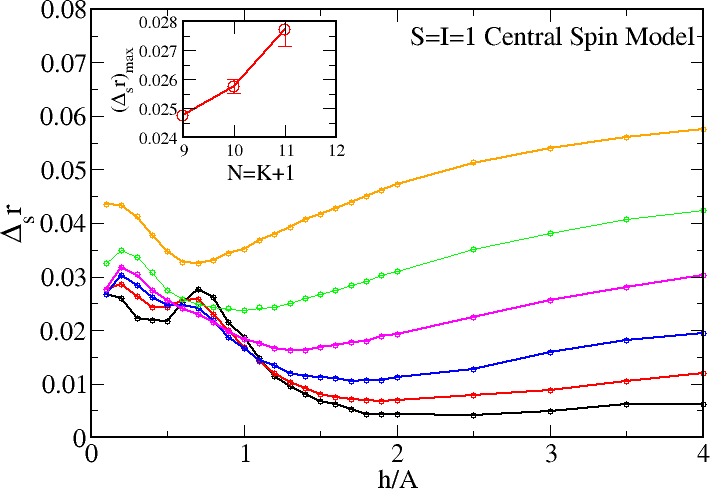}
  \caption{Top panels: The expectation value $\langle r\rangle_p$ as
    a function of disorder strength in central spin modelss of spin
    length $S=I=1/2$ (left) and $S=I=1$ (right)
    for various system sizes and pertaining numbers of disorder realizations.
    The error bars are determined by Eqs.~(\ref{stat13}), (\ref{stat14}). 
    Inset: The standard deviation $\Delta_pr$ as a function of disorder
    strength. The horizontal lines indicate the expected values
    for GOE and Poissonian statistics as given in
    Eqs.~(\ref{ratgap4})-(\ref{ratgap10}).\\
    Bottom panels: The standard deviation $\Delta_sr$ according to
    Eq.~(\ref{stat14}) for the same parameters as in the top panels.
    For a more detailed view on the data at large system sizes see
    also Fig.~\ref{censpinfig3}.
    The insets show the maximum of the standard deviation as a function of
    system size.}
  \label{censpinfig2}
\end{figure*}

Fig.~\ref{censpinfig2} displays data for the central spin
model analogous to Fig.~\ref{chainfig2} for the
Heisenberg chain. Comparing the top panels of both figures suggests
that a transition from the (approximately) ergodic to the many-body localized
phase at the inflection point  $h=h_{\rm inf}\lesssim 1$, which is consistent
with the findings of Ref.~\cite{Hetterich18} for $A,h\gg J$.
Here the fact that the transition occurs at
about the same disorder strength for different system sizes depends on the
scaling factor $1/K$ in front of the second term in the Hamiltonian
(\ref{hamiltonian}).
Also the sample-to-sample standard deviation $\Delta_sr$ plotted in the bottom
panels of Fig.~\ref{censpinfig2} behaves similarly as for the Heisenberg chain:
For large enough systems sizes this quantity developes a maximum
$(\Delta_sr)_{\rm max}$ near $h=h_{\rm inf}$
whose value increases monotonously with system size, as shown in the insets.
Thus, we have qualitatively the same situation as for the Heisenberg chain.
\begin{figure}
  \includegraphics[width=0.8\columnwidth]{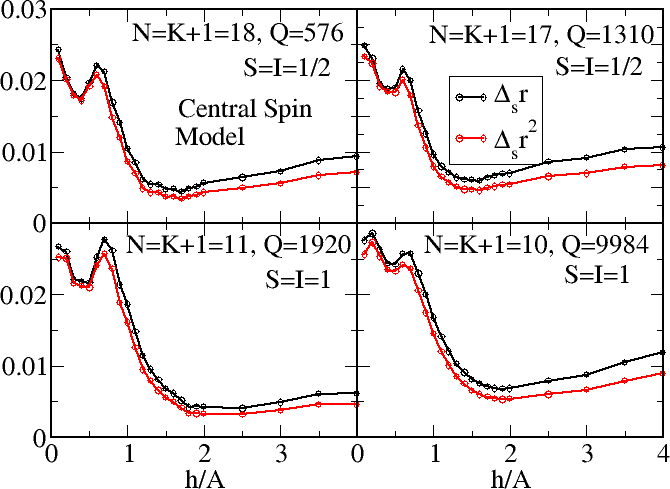}
  \caption{The sample-to-sample standard deviation $\Delta_sr$ along with the
    square root of ``variance of the variance'' (\ref{stat29}) as a function of
    disorder strength $h$ for central spin models of spin length $S=I=1/2$
    (top panels) and $S=I=1$ (bottom panels).}
  \label{censpinfig3}
\end{figure}

Moreover, as seen in Fig.~\ref{censpinfig3}, the square root of the 
``variance of the variance'' (\ref{stat29}) follows, similarly as for the
Heisenberg chain, closely the standard deviation $\Delta_sr$ 
as a function of disorder strength for both spin lengths $I=1/2$ and $I=1$.
\begin{figure*}[!p]
  \includegraphics[width=0.8\columnwidth]{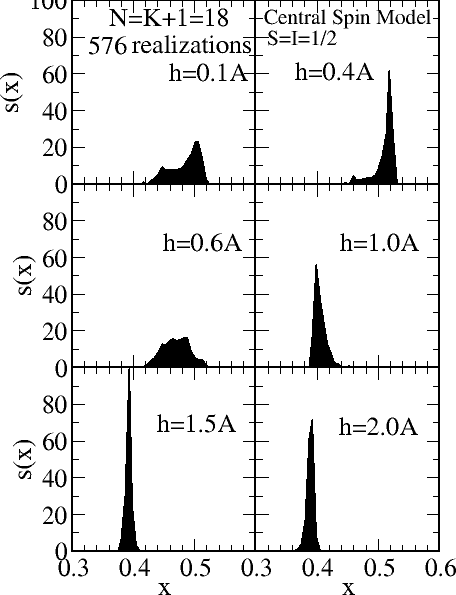}
  \includegraphics[width=0.8\columnwidth]{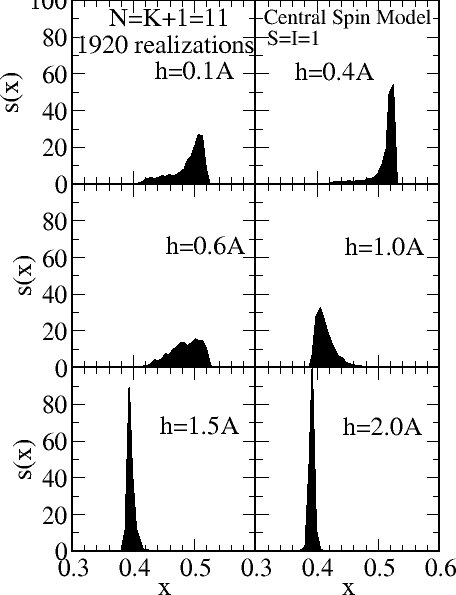}
  \caption{The probability distribution (\ref{stat6}) for the
    realization-specific average $\langle r\rangle_{\alpha}$
    for disordered cenral spin systems of spin length $S=I=1/2$ (left) and
    $S=I=1$ (right).
    The disorder strengths $h$ in both panels are the same as in
    Fig.~\ref{censpinfig1}.}
  \label{censpinfig4}
\end{figure*}
This is consistent with the probability distribution
(\ref{stat6}) for the realization-specific average $\langle r\rangle_{\alpha}$
shown in Fig.~\ref{censpinfig4}.
As already seen in the case of the Heisenberg chain, the probability
distribution is much
narrower than the distribution $p(r)$ and
becomes significantly broader in the transition region.

\section{Summary and Outlook}
\label{SummaryandOutlook}

We have compared the transitions between ergodic and many-body localized phases
in disodered Heisenberg chains as well as central spin models
composed of spins of length $1/2$ and $1$.
A useful new tool we introduce is the sample-to-sample standard
deviation $\Delta_sr$ 
of the expectation value $\langle r\rangle_{\alpha}$ of the consecutive-gap
ratio in an individual disorder realization (sample) $\alpha$.
This quantity assumes, for both types of systems and spin lengths, a maximum
as a function of disorder strength, accompanied by an inflection point of
$\langle r\rangle$. These are typical features of a phase transition
where the latter quantity play the role of an order parameter.
The critical disorder strength deduced from these observations turn out to be
smaller than those reported in the recent literature.

Further information about the transitions is contained in the 
probability distribution of the expectation values within a
given disorder realization. We expect the study of this probability
distribution and its moments to be a useful tool in the investgation
of phenomena related to many-body localization also in other systems.

\acknowledgments

We thank F. Evers and F. G\"ohmann for useful discussions,
and M. Trivelato for collaboration on an earlier stage of this project.
J.S. acknowledges support by FAPESP and the hospitaliy of the University of
Sao Paulo at Sao Carlos and of IIP Natal.
J.C.E. acknowledges support from the Sao Paulo Research Foundation (FAPESP)
Grants No. 2016/08468-0, No. 2018/19017-4, No. 2020/00841-9, and and from
Conselho Nacional de Pesquisas (CNPq), Grant No. 306122/2018-9.

\end{document}